# PROCESSING STUDIES OF X-BAND ACCELERATOR STRUCTURES AT THE NLCTA*

C. Adolphsen, W. Baumgartner, K. Jobe, F. Le Pimpec, R. Loewen, D. McCormick,
M. Ross, T. Smith, J.W. Wang, SLAC, Stanford, CA 94309 USA
T. Higo, KEK, Tskuba, Ibaraki, Japan

*Abstract*

RF processing studies of 1.8-m X-band (11.4 GHz) traveling wave structures at the Next Linear Collider Test Accelerator (NLCTA) have revealed breakdown-related damage at gradients lower than expected from earlier tests with standing wave and shorter, lower group velocity traveling wave structures. To understand this difference, a series of structures with different group velocities and lengths are being processed. In parallel, efforts are being made to improve processing procedures and to reduce structure contaminants and absorbed gases. This paper presents results from these studies.

## 1 INTRODUCTION

During the past eight years, several prototype NLC/JLC X-band accelerator structures were built to test methods of long-range transverse wakefield suppression and to study high gradient performance [1]. Most of the structures are 1.8-m long (206 cells), traveling wave ($2\pi/3$ phase advance per cell) and nearly constant gradient. To achieve the last condition, the group velocity relative to c ($\equiv v_g$) varies from 12% to 3% along the 1.8-m structures, yielding a 100 ns fill time. Four of these structures were installed in the NLCTA to operate initially at 50 MV/m unloaded gradient with 240-ns pulses, the first-stage NLC rf parameters at the time. These parameters have since evolved to 70 MV/m and 386 ns to reduce the NLC cost. To achieve this higher gradient in the 1.8-m prototypes, about 170 MW of input power per structure is required.

During the past year, the high power testing capability at the NLCTA increased significantly, allowing automated, around-the-clock processing at higher gradients. During this period, it was discovered that the net rf phase advance through the prototype structures had increased by roughly 20 degrees per 1000 hours of operation at gradients as low as 50 MV/m [2]. This was surprising since earlier tests had shown that gradients of more than 80 MV/m could be readily achieved in standing wave and shorter, lower $v_g$ traveling wave structures [3]. Such structures had been used because high gradients could be achieved with the limited rf power available at the time. A major clue as to the cause of this discrepancy was that most damage in the 1.8-m structures occurred in the upstream end where $v_g$ is highest (12% to 5%). Little damage occurred in the downstream end where the $v_g$ is comparable to that in the early test structures (< 5%).

## 2 RF CIRCUIT AND BREAKDOWN

From the 1.8-m and early structure results, it was hypothesized that the upstream damage was related to the higher group velocity in this region. This relation was proposed in part because the rf power required to achieve a given gradient increases with $v_g$. Also, if the structure is viewed as a transmission line and rf breakdown as a load impedance, the fraction of incident power absorbed during breakdown increases with $v_g$. The net effect, assuming the breakdown impedance is real and small compared to the structure impedance, is that the power absorbed, $P_{abs}$, scales as,

$$P_{abs} \sim \frac{v_g^2}{(R/Q)^2} \frac{\sin(\varphi)}{\varphi \sin(\varphi) + 2 v_g \cos(\varphi)} \text{Gradient}^2$$

where R is the cell shunt impedance, Q is the cell quality factor, and $\varphi$ is the phase advance per cell. Only $v_g$ varies significantly along the 1.8-m structures, so if damage is directly related to absorbed power, the gradient at which a given level of damage occurs scales as the inverse of $v_g$.

Another possible explanation for the greater damage upstream is that reflected rf energy from a breakdown causes an over-voltage upstream that initiates further breakdown within the same rf pulse. Such events have been observed, although more than 70% of breakdowns appear to occur in a single location.

## 3 STRUCTURE COMPARISONS

To better understand high gradient limits, a series of structures with different lengths and $v_g$ profiles are being processed. In addition, various improvements are being made to the structure cleaning and handling procedures. In particular, a degassing process has been adopted that includes 'wet' and 'dry' $H_2$ firing at 950 ºC, a two-week vacuum bake at 650 ºC, and a one-week in-situ bake at 220 ºC. This procedure was used for the last two low $v_g$ structure tests described below. Finally, the rf processing protocol has been improved in an attempt to minimize structure damage. Processing is done at a 60 Hz pulse rate under computer control. Typically a target gradient is achieved using a sequence of progressively longer pulses (e.g., 50, 100, 170 and 240 ns). If breakdown is detected (> 10% drop in transmitted energy), the rf is shut off, brought back to full amplitude at a reduced pulse width (50 ns), then ramped to full pulse width. The recovery time is about one minute.

---

* Work Supported by DOE Contract DE-AC03-76F00515.

The first structure tested in this program was made by cutting off the last 52 cells of one of the 1.8-m structures (DS2). This region was chosen because it showed no discernible phase shift from its previous operation at less than 50 MV/m. The shortened structure was operated for about 1700 hours with 240 ns square pulses at gradients up to 73 MV/m. The structure incurred some damage during initial processing to over 70 MV/m. This was indicated by a 5° net phase shift as measured using beam-induced rf. However, subsequent running for 650 hours at 65 MV/m and 750 hours at 70 MV/m produced no measurable phase change (< 1°). The breakdown rates at the end of the run the were fairly low: about one per hour at 70 MV/m and one per four hours at 65 MV/m. The value at 65 MV/m would be marginally acceptable for the NLC when scaled to the proposed operating conditions.

The goal of the next test was to compare the performance of two structures of different lengths but the same $v_g$ profile. For this purpose, two new structures were fabricated with an initial $v_g$ of 5%. One was 20 cm long (23 cells) and the other 105 cm long (120 cells) with the first 23 cells identical to those in the shorter structure. The cell parameters were varied along the structures to make the iris surface fields constant, thus eliminating field strength differences as a factor in the comparison. Also, power was split evenly to feed the structures so they witnessed the same fields. Unfortunately, an unexpectedly large reflection from the loads on the shorter structure produced a roughly 10% over-voltage within it, so a simple comparison of the results cannot be made. Nonetheless, the breakdown rates in the two structures were comparable during processing to the highest gradients, so the short structure was not significantly superior in this regard. In general, these rates were similar to those measured in the previous test but the phase shift was less (3.5°) after 500 hours of operation.

During the two tests discussed above, one of the 1.8-m prototype structures that had not been previously processed (DDS3) was operated to determine the gradient at which damage began. Although a 'more gentle' processing protocol was used than in the past, a damage threshold of 45-50 MV/m was found, consistent with the earlier 1.8-m structure results. This range is markedly lower than the 70-75 MV/m level for the lower $v_g$ structures. Also, at the various pulse lengths, the gradient at which the breakdown rate initially exceeded several per hour occurred at the 50-60 MV/m level in the low $v_g$ structures compared to 30-40 MV/m in the 1.8-m structure.

Currently, two 53-cm structures are being processed in parallel, one with a 5.0-3.3% $v_g$ profile, and one with a 3.3-1.6% $v_g$ profile. An in-line stainless-steel load is being used to reduce the power to the lower $v_g$ structure. The goal was to equalize the gradients, but with only one iteration of the load design, the resulting attenuation is less than desired, yielding a 6% higher gradient in the lower $v_g$ structure.

This pair of structures has processed much more rapidly than ones in the past, so far to 77 MV/m for the lower $v_g$ structure and 73 MV/m for the higher $v_g$ structure. No discernible phase shift has been observed in either structure after achieving these levels. After 280 hours of operation, the breakdown rate in the lower $v_g$ structure was about 1 per hour at 70 MV/m; in the higher $v_g$ structure, the rate was about one per two hours at 66 MV/m. The faster processing may be related to the deeper etching (few microns versus sub-micron) used in cleaning the cells before assembly, although refinements of the new baking procedures may have contributed as well. The current program is to process these structures to higher gradients to find their damage threshold.

## 4 BREAKDOWN OBSERVATIONS

Various studies have been done to learn more about breakdown, including the examination of cell irises after processing, the determination of breakdown rate dependence on pulse length and gradient, and the measurement of various emission signals (rf, light, X-rays, sound [4], vacuum pressure and electron currents) before, during, and after breakdown. No consistent precursors to breakdown have been found nor is there definitive evidence from SEM, EDX [5] and RGA measurements as to what triggers breakdown. The breakdown rate has little dependence on the structure vacuum pressure for values up to $10^{-7}$ torr. The rate decreases slowly once a given gradient has been reached; a faster reduction can be achieved by first processing to a higher gradient. In general, the breakdown rate rises exponentially with gradient, and more than linearly with pulse length.

The most detailed information on breakdown has come from analyzing the transmitted and reflected rf from the structures. Figure 1 shows an example. Such information can be used to determine the breakdown time during the pulse, the breakdown location in the structure and the amount of rf power absorbed after breakdown. In general, the transmitted power falls to essentially zero about 100 ns after breakdown. It fully recovers several microseconds after the main rf pulse, when the power from the continued discharge of the NLCTA pulse compression system is many orders of magnitude smaller. The distribution of peak reflected power is broad and peaked at low values: the mean is typically 10% to 20% of the input power. Up to about 80% of the incident rf energy is absorbed after breakdown begins, although only a very small fraction ($10^{-5}$) of the rf energy, if converted to heat, is needed to produce a pit of the size observed on the cell irises. Simulations using particle-in-cell codes of the rf interaction with breakdown-generated electrons and ions are being done to understand this energy absorption phenomenon [6].

The distribution of breakdown times during the pulse is fairly uniform. One puzzling result is that about 10% of the breakdowns occur after the main pulse, when the power is less than 10% of maximum. The distribution of breakdown locations in the structure is generally skewed toward the upstream end. It differs depending on whether the structure is being processed to a higher gradient for the first time or it is running at a gradient lower than has been achieved. Figure 2 shows a plot of fractional missing energy versus breakdown location in each case. A common feature is the clustering of events near the input coupler. At the highest gradients, breakdown occurs deeper into the structure, mostly with high missing energy (> 40%). The events near the coupler include a distinct set of 'soft' events with lower missing energy, slower falloff in transmitted power, and a common iris origin (probably the input cell iris) as evidenced by the phase of the reflected rf. A new coupler design that increases group velocity in a two-step transition from the input rectangular waveguide to the structure cells will be tested shortly in an attempt to eliminate the events near the coupler.

Perhaps the most important observation from the rf measurements is that for the longest pulse, highest gradient operation, breakdown tends to occur clustered in time in localized regions of the structure (so called 'spitfests'). Figure 3 shows an example of the breakdown location history when operating in this regime. In general, the breakdown location propagates to the upstream end of the structure, where the breakdowns are mainly 'soft' and occur at a low rate. This migration process tends to repeat after a large-missing-energy breakdown occurs deeper in the structure. Such cycles suggest that breakdown can cause secondary damage that lowers the threshold on subsequent pulses. Evidence for secondary damage is seen on the irises after processing in that the number of pits is much larger (10-100 times) than the number of rf trips that had occurred during processing. Also, the occurrence of spitfests increases with pulse length, and the distribution of breakdown times during the pulse is surprisingly flat after the pulse has been lengthened. If such collateral damage depends on group velocity, it would explain why cell irises with the same surface field both damage and process differently in low and high group velocity structures.

## 5 OUTLOOK

The results from the low $v_g$ structure tests are encouraging for the NLC goal of 70 MV/m unloaded gradient operation. However, the average iris radius of the current test structures is too small to meet NLC short-range wakefield requirements. To increase the iris size while maintaining low $v_g$, a structure with higher phase advance per cell (150° instead of 120°) will be used [7]. Three designs are being developed, one with 5.1% initial $v_g$ that is 0.9 m long and two with 3.2% initial $v_g$ that are 0.9 m and 0.6 m long. Prototypes of these designs are scheduled to be processed by the end of the year. Also, tests of standing wave structures have started and will continue to fully evaluate their feasibility [8]. Finally, various methods of pre-treating the structures will be tried, including glow discharge cleaning.

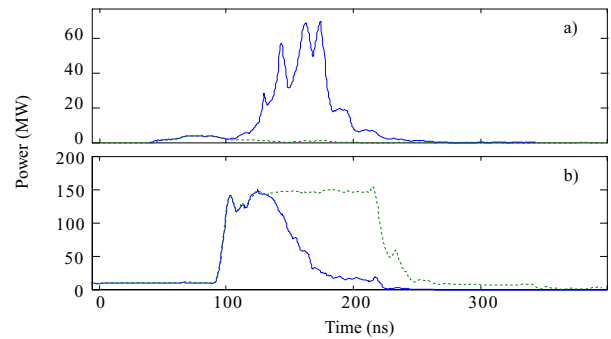

Fig. 1: Example of (a) reflected and (b) transmitted power for a breakdown pulse (solid) and normal pulse (dashed).

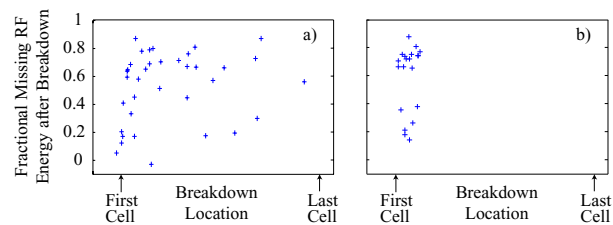

Fig. 2: Fractional missing energy versus breakdown location when (a) processing to a higher gradient and (b) running at a gradient lower than the maxiumum achieved.

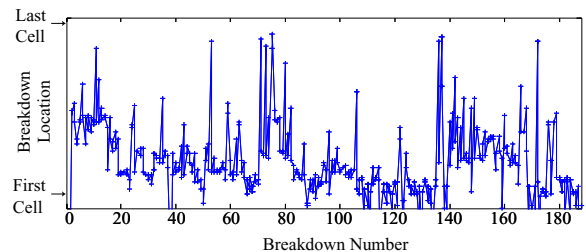

Fig. 3: Sequence of breakdown locations during 'spitfests.'